%Paper: hep-th/9406051
%From: LESNIEWSKI@huhept.harvard.edu
%Date: Wed, 8 Jun 1994 21:54:29 -0400 (EDT)

\newbox\leftpage \newdimen\fullhsize \newdimen\hstitle \newdimen\hsbody
\tolerance=10000\hfuzz=10000pt
%    tolerance=1000 \hfuzz=20pt
\def\printertype{ps: }% default postscript
\def\qms{\def\printertype{qms: }% qms lasergrafix
\ifx\answ\bigans\else\voffset=-.4truein\hoffset=.125truein\fi}
\def\bigans{b }
\message{ big or little (b/l)? }\read-1 to\answ
\ifx\answ\bigans\message{(This will come out unreduced.}
\magnification=1200\baselineskip=14pt plus 2pt minus 1pt
\hsbody=\hsize \hstitle=\hsize %take default values for unreduced format
\else\message{(This will be reduced.} \let\lr=L
\magnification=1000\baselineskip=16pt plus 2pt minus 1pt
\voffset=-.31truein\vsize=7truein\hoffset=-.59truein% apple lw
\hstitle=8truein\hsbody=4.75truein\fullhsize=10truein\hsize=\hsbody
\output={\ifnum\pageno=0 %%% This is the HUTP version
  \shipout\vbox{\special{\printertype landscape}\makeheadline
    \hbox to \fullhsize{\hfill\pagebody\hfill}}\advancepageno
  \else
  \almostshipout{\leftline{\vbox{\pagebody\makefootline}}}\advancepageno
  \fi}
\def\almostshipout#1{\if L\lr \count1=1 \message{[\the\count0.\the\count1]}
      \global\setbox\leftpage=#1 \global\let\lr=R
  \else \count1=2
    \shipout\vbox{\special{\printertype landscape}
      \hbox to\fullhsize{\box\leftpage\hfil#1}}  \global\let\lr=L\fi}
\fi

\input mssymb.tex
\font\bigfont=cmr10 scaled\magstep3

\def\section#1#2{\vskip32pt plus4pt \goodbreak \noindent{\bf#1. #2}
	\xdef\currentsec{#1} \global\eqnum=0 \global\thmnum=0}

\newcount\thmnum
\global\thmnum=0
\def\prop#1#2{\global\advance\thmnum by 1
	\xdef#1{Proposition \currentsec.\the\thmnum}
	\bigbreak\noindent{\bf Proposition \currentsec.\the\thmnum.}
	{\it#2} }
\def\define#1#2{\global\advance\thmnum by 1
	\xdef#1{Definition \currentsec.\the\thmnum}
	\bigbreak\noindent{\bf Definition \currentsec.\the\thmnum.}
	{\it#2} }
\def\lemma#1#2{\global\advance\thmnum by 1
	\xdef#1{Lemma \currentsec.\the\thmnum}
	\bigbreak\noindent{\bf Lemma \currentsec.\the\thmnum.}
	{\it#2}}
\def\thm#1#2{\global\advance\thmnum by 1
	\xdef#1{Theorem \currentsec.\the\thmnum}
	\bigbreak\noindent{\bf Theorem \currentsec.\the\thmnum.}
	{\it#2} }
\def\cor#1#2{\global\advance\thmnum by 1
	\xdef#1{Corollary \currentsec.\the\thmnum}
	\bigbreak\noindent{\bf Corollary \currentsec.\the\thmnum.}
	{\it#2} }

\newcount\eqnum
\global\eqnum=0
\def\num{\global\advance\eqnum by 1
	\eqno({\rm\currentsec}.\the\eqnum)}
\def\eqalignnum{\global\advance\eqnum by 1
	({\rm\currentsec}.\the\eqnum)}
\def\ref#1{\num  \xdef#1{(\currentsec.\the\eqnum)}}
\def\eqalignref#1{\eqalignnum  \xdef#1{(\currentsec.\the\eqnum)}}

\def\title#1{\centerline{\bf\bigfont#1}}

\newcount\subnum
\def\Alph#1{\ifcase#1\or A\or B\or C\or D\or E\or F\or G\or H\fi}
\def\subsec{\global\advance\subnum by 1
	\vskip12pt plus4pt \goodbreak \noindent
	{\bf \currentsec.\Alph\subnum.}  }
\def\newsubsec{\global\subnum=1 \vskip6pt\noindent
	{\bf \currentsec.\Alph\subnum.}  }
\def\today{\ifcase\month\or January\or February\or March\or
	April\or May\or June\or July\or August\or September\or
	October\or November\or December\fi\space\number\day,
	\number\year}

\def\intsec{I}
\def\fredsec{II}
\def\suisec{III}
\def\suiisec{IV}
\def\svsec{V}
\def\bC{\Bbb{C}}
\def\bR{\Bbb{R}}
\def\bZ{\Bbb{Z}}
\def\su{{\cal M}}
\def\sv{{\cal X}}
\def\dmu#1{d\mu_r(#1)}
\def\rker#1#2{K^r(#1,#2)}

\def\rkerv#1#2{K^r\ssub V(#1,#2)}

\def\ssub#1{_{\scriptscriptstyle #1}}
\def\Tr{\mathop{\rm Tr}\nolimits}

\def\del{\partial}

\def\Str{\mathop{\rm Str}\nolimits}
\def\ol{\overline}
\def\smooth#1{C^\infty(#1)}
\def\norm#1{\Vert #1 \Vert}
\def\inv{^{-1}}
\def\hil{{\cal H}}
\def\Hru{{\cal H}_r(\su)}
\def\hru{{\cal H}_r(D)}
\def\hrv{{\cal H}_r(V)}
\def\hrsv{{\cal H}_r(\sv)}
\def\trd{{\cal T}_r(D_{m,n})}
\def\tru{{\cal T}_r(\su)}
\def\trv{{\cal T}_r(V)}
\def\trsv{{\cal T}_r(\sv)}
\def\calp{{\cal P}}
\def\apoly{{\cal A}^{pol}}
\def\inq{I_{n|q}}
\def\star{^\ast}
\def\cald{{\cal D}}

{\baselineskip=12pt
\nopagenumbers
\line{\hfill \bf HUTMP B331}
\line{\hfill \bf \today}
\vfill
\title{Supersymmetry and Fredholm Modules}
\medskip
\title{Over Quantized Spaces}
\vskip1in
\centerline{{\bf David Borthwick}$\star$, {\bf Slawomir
Klimek}$^{**}$
\footnote{$^1$}{Supported in part by the National Science
Foundation under
grant DMS--9206936},}
\vskip.3cm
\centerline{{\bf Andrzej Lesniewski}$\star$\footnote{$^2$}
{Supported in part by the Department of Energy under grant
DE--FG02--88ER25065}, and {\bf Maurizio
Rinaldi}$\star$\footnote{$^3$}
{Supported in part by the Consiglio Nazionale delle Ricerche
(CNR)}}
\vskip12pt
\centerline{ $\star$Lyman Laboratory of Physics}
\centerline{Harvard University}
\centerline{Cambridge, MA 02138, USA}
\vskip12pt
\centerline{ $^{**}$Department of Mathematics}
\centerline{IUPUI}
\centerline{Indianapolis, IN 46205, USA}
\vskip1in\noindent
{\bf Abstract.}
The purpose of this paper is to apply the framework of non-commutative
differential geometry to quantum deformations
of a class of K\"ahler manifolds.  For the examples of the
Cartan domains of type I and flat space, we construct Fredholm
modules over the quantized manifolds using the supercharges which
arise in the quantization of supersymmetric generalizations of
the manifolds.  We compute an explicit formula for the Chern character
on generators of the Toeplitz $\bC^\ast$-algebra.
\vfill\eject}

\section\intsec{Introduction}

\newsubsec
Since the early work on quantum mechanics ([15], [3], [12], [4]) by
Heisenberg, Born, Jordan, and Dirac, it has been generally
recognized
that ordinary geometry does not apply to the subatomic world. In
order to describe the physical phenomena in that world, the classical
notion of phase space needs to be replaced by a non-commutative
algebra of ``quantum observables".   The coordinates $p$ and $q$ on
the phase space $\bR^2$ are replaced by
generators $p$ and $q$ that obey the famous commutation relation
$[q,p]=i\hbar$. This ``quantization'' procedure amounts to studying
a non-commutative deformation of a flat space, and,
from a geometric viewpoint quantum mechanics emerges as some
form of symplectic geometry on this non-commutative space. The
classical algebra of functions on phase space arises as the
$\hbar\rightarrow 0$
limit of the deformed algebra, and the Poisson bracket of two
observables
turns out to be the subleading term in the small $\hbar$ expansion of
the commutator of the corresponding quantized observables. Much
work has been done since the early quantum mechanics
on extending this procedure to more general,
non-flat phase spaces, resulting in powerful theories known as
geometric quantization, deformation quantization, quantum groups,
etc.

\subsec
In the mid-eighties A. Connes [9] proposed a general scheme of
non-commutative differential geometry which is ideally suited
to describe the geometry of quantum theory. The central concept
of this scheme is a non-commutative $\bC\star$-algebra $\cal A$
which plays the role of an algebra of continuous functions on the
(putative) non-commutative space. The de~Rham cohomology of a smooth
manifold is replaced by the cyclic cohomology of $\cal A$, while
elements of the group $K_0({\cal A})$ play the role of vector
bundles over the non-commutative space. Another central concept of
Connes' framework is that of a $p$-summable Fredholm module over
$\cal A$, which replaces the classical notion of metric structure
on a manifold. This is defined as a triple $({\cal H},\rho , Q)$,
where $\cal H$ is a $\bZ_2$-graded Hilbert space, $\rho$ is an
action of $\cal A$ by bounded operators on $\cal H$, and $Q$ is
a self-adjoint operator on $\cal H$ which is odd with respect to
the $\bZ_2$-grading (see Section II for a precise definition).
In examples, the operator $Q$ is often a Dirac type operator.
To a Fredholm module over $\cal A$, Connes associates a fundamental
cocycle in cyclic cohomology, called the Chern character.

\subsec
A conceptual framework of quantization which fits the scheme of
non-commuative differential geometry was proposed by Rieffel in [19].
This framework relies on the use of $\bC\star$-algebras and precise
operator norm estimates (rather than formal power series in Planck's
constant), and we refer to it as non-perturbative deformation
quantization. Examples of quantized spaces studied within this
framework include quantized tori (see [20] for a review and references)
and quantized flat spaces (see [8] for some recent results and
references). In [17], [7], [5], and [6] we studied quantum deformations
of a class of hermitian symmetric spaces and superspaces, namely
the Cartan domains and superdomains. In each case, we constructed a
family of $\bC\star$-algebras of ``quantized functions'' and verified
that these $\bC\star$-algebras are indeed quantum deformations of the
corresponding classical algebras of functions. In this paper and
its sequel we study Fredholm modules over these algebras and the
associated Chern characters.

\subsec
As explained by Witten in his work on supersymmetry [21], it is
natural to regard Dirac type operators as generators of supersymmetries
(``supercharges'') in certain physical systems involving bosons
and fermions. This suggests that a natural way of constructing
Fredholm modules over a quantized manifold is first to quantize a
supersymmetric generalization of the manifold (see e.g. [2] for an
introduction to the theory of supermanifolds), and then to take $Q$ to be
a supercharge generating the supersymmetry. Following this idea,
we construct Fredholm modules over the quantized type I Cartan
domains and over quantized flat spaces. Our construction relies on
[5], where we studied the relevant supersymmetric theories. We
introduced there the notion of a super Toeplitz operator and the
$\bC\star$-algebra
generated by such operators. The Fredholm modules we construct
bear a certain resemblance to those constructed for the Toeplitz
algebra over the circle in Section 4.2 of [11].

\subsec
The paper is organized as follows. In Section II we briefly review
Connes' formalism of non-commutative differential geometry. In
Sections III and IV we study Fredholm modules over the quantized
type I Cartan domains, and in Section V we study Fredholm modules
over quantized flat spaces.

\section\fredsec{Fredholm Modules and Their Chern Characters}

\newsubsec
In this section we briefly review the notion of a $p$-summable
Fredholm module over a $\bC\star$-algebra $\cal A$. From a physical
point of view, the concept of a Fredholm module captures the essential
features of a quantum supersymmetric system: the $\bC\star$-algebra
$\cal A$ is the algebra of observables, the Dirac operator is the
supersymmetry generator, and its square is the Hamiltonian of the
system.

Let ${\cal A}$ be a trivially $\bZ_2$-graded $\bC\star$-algebra (all
elements are even).
Recall ([9], [11]) that a $p$-summable Fredholm module over ${\cal A}$
is a triple $({\cal H}, \rho , Q)$ such that:

\noindent
(i) ${\cal H}$ is a $\bZ_2$-graded Hilbert space. We denote by $\Gamma$
the grading operator and by ${\cal H}_0$ and ${\cal H}_1$ the
homogeneous subspaces of ${\cal H}$.

\noindent
(ii) $\rho:{\cal A}\longrightarrow{\cal L}({\cal H})$ is a grading
preserving
$\ast$-homomorphism of ${\cal A}$ into the algebra of bounded linear
operators on ${\cal H}$. For notational simplicity, we will suppress
$\rho$ in all formulas throughout the rest of this paper.

\noindent
(iii) $Q$ is a self-adjoint operator on ${\cal H}$ which is odd under
the $\bZ_2$-grading, i.e.
$$
Q\Gamma +\Gamma Q=0\; ,\num
$$
and such that for any $\epsilon > 0$,
$$
(Q^2+I)^{-1/2}\in I_{p+\epsilon}({\cal H})\; .\ref{\psumref}
$$
Here, $I_p({\cal H})$ denotes the $p$--th Schatten class of operators
on ${\cal H}$. It is natural to regard the smallest number $p$ in
\psumref\ as the dimension of the non-commutative space.

\noindent
(iv) The subalgebra ${\cal A}^Q$ consisting of those $a\in\cal A$
for which the commutator $[Q,a]$ is bounded is dense in $\cal A$.

We refer to $Q$ as the Dirac operator. In the following, we will
denote $H:=Q^2$ and refer to it as the Laplace operator. Clearly,
$H$ is a positive self-adjoint operator.

\subsec
A $p$-summable Fredholm module defines a fundamental cocycle in the cyclic
cohomology of ${\cal A}^Q$, called the Chern character [9]. In this paper
we use the Chern character of [16], which is a cocycle in the entire
cyclic cohomology [10] of ${\cal A}^Q$. It has the advantage of being
expressed in terms of the heat kernel of the Laplace operator (very
much like the McKean-Singer formula in index theory), which leads to
useful integral representations. Its truncation to cyclic cohomology
is discussed in [11]. This cocycle, ${\rm Ch}^\beta(Q)=\{{\rm Ch}
^\beta_{2k}(Q)\}_{k=0}^\infty$, where each ${\rm Ch}^\beta_{2k}(Q)$
is a $(2k+1)$-linear functional on ${\cal A}^Q$, is defined as follows.
For $a\in{\cal L}({\cal H})$ and $t>0$ we define the unbounded, densely
defined operator
$$
a(t):=e^{-tH}ae^{tH}\; .\num
$$
For $\beta >0$ and $a_0,a_1\dots ,\; , a_{2k}\in {\cal A}^Q$ we set
$$
{\rm Ch}^\beta_{2k}(Q)(a_0,a_1,\dots ,\; a_{2k}):=\beta^{-k}
\int_{\sigma^\beta_{2k}}\Str\Bigl\{a_0 \> [Q,a_1](t_1)\dots
[Q,a_{2k}]
(t_{2k})\> e^{-\beta H}\Bigr\}\; d^{2k}t\; ,\ref{\cocycle}
$$
where $\sigma^\beta_{n}:=\{(t_1,\dots ,t_n)\in{\bR}^n:\;
0\leq t_1 \leq\dots\leq t_n\leq\beta\}$, and where
$\Str$ denotes the supertrace,
$$
\Str (A):=\Tr(\Gamma A)\; .\num
$$
The key analytic input ensuring the existence of \cocycle\ is the
following inequality [16]. For $s_j\geq 0$, $j=0,\ldots,n$, with
$\sum_{j=0}^n s_j=\beta$, and $A_0, A_1,\dots ,A_n\in{\cal L}
({\cal H})$,
$$
\Bigl|\Tr(A_0e^{-s_0 H}A_1e^{-s_1H}\ldots A_n e^{-
s_nH})\Bigr|\leq \prod_{j=0}^n
\Vert A_j\Vert\; \Tr(e^{-\beta H})\ref{\jlo}
$$
(we use this estimate with $s_0=t_1,\; s_1=t_2-t_1,\dots\; ,s_n=
\beta-t_n$).  The inequality \jlo\ is proven using H\"older's
inequality and holds for any positive operator in place of $H$.

\subsec
For the case of Fredholm modules arising from the quantized
K\"ahler supermanifolds discussed later in this paper,
we will have two Dirac operators $Q_1$ and $Q_2$ which generate
an $N=2$ supersymmetry algebra, namely
$$
\eqalign{
&Q_1^2 = Q_2^2 = H,\cr
&Q_1Q_2+Q_2Q_1=0\; .\cr}\num
$$
These two operators will be essentially
self-adjoint on some dense subspace $\calp \subset {\cal H}$
and will have the structure:
$$
\eqalign{&Q_1 = d + d^\ast, \cr
&Q_2 = i(d - d^\ast), \cr}  \ref{\qstrc}
$$
where $d$ is an operator such that $d^2 = 0$, Dom$(d) = \calp$.
We define ${\cal A}^1:={\cal A}^{Q_1}\cap{\cal A}^{Q_2}$.  Here we make an
important assumption that ${\cal A}^1$ be dense in ${\cal A}$ (this
is the key element which limits the generality of the following
theorem). This will clearly be the case for the examples we consider.

\thm\homthm{Let $({\cal H},\rho, Q_1)$ and $({\cal H},\rho,
Q_2)$ be Fredholm modules over ${\cal A}$, with $Q_1$ and $Q_2$ as in
\qstrc.
Then the corresponding Chern characters define the same cohomology
class in the entire cyclic cohomology of ${\cal A}^1$.}
\medskip\noindent{\it Proof.}
The theorem is proven through a homotopy argument.  We can set
$\beta = 1$
with no loss of generality.  We form the family of operators
$$
Q(\lambda) := Q_1 \cos\lambda + Q_2 \sin\lambda
= e^{i\lambda}d + e^{-i\lambda}d^\ast,  \num
$$
for $0\le \lambda < 2\pi$, which interpolates between $Q_1$ and
$Q_2$: $Q(0)=Q_1$, $Q(\pi/2)=Q_2$. Clearly, $Q(\lambda)$ is
essentially self-adjoint on $\calp$, and the commutator of $Q(\lambda)$
with any element of ${\cal A}^1$ is bounded. We proceed as in [13] to show
that ${d\over d\lambda}{\rm Ch}(Q(\lambda))$ exists and is equal
to a coboundary, so that ${\rm Ch}(Q_1)$ and ${\rm Ch}(Q_2)$
are cohomologous. Note that the technical assumptions (i) and (ii) of
Section III of [13] are not satisfied in our case and we need to make
some changes in the argument.

Observe that the $Q(\lambda)$ obey the following algebra:
$$
[Q(\lambda),Q(\mu)] = 2 \cos(\lambda-\mu)\; H.  \num
$$
In particular, $Q(\lambda)^2=H$, and so $H$ is the Laplace
operator
corresponding to $Q(\lambda)$ for any $\lambda$.  On $\calp$
we can take the derivative
$$
Q'(\lambda) := {d\over d\lambda} Q(\lambda)
= - Q_1 \sin\lambda + Q_2\cos\lambda = Q(\lambda + \pi/2),
\ref{\qprref}
$$
and we thus see the $Q'(\lambda)$ is also essentially self-adjoint on
$\calp$.
Define $G^\lambda = \{G^\lambda_{2k-1}\}_{k=1}^\infty$, where
each
$G^\lambda_{2k-1}$ is a $2k$-linear functional on ${\cal A}^1$
given by
$$
\eqalignno{&G_{2k-1}^\lambda(a_0, \ldots, a_{2k-1})
&\eqalignref{\gdef} \cr
&:=\sum_{l=0}^{2k-1} (-1)^{l+1}
\int_{\sigma^1_{2k}} \Str \Bigl\{ a_0 \>[Q(\lambda),a_1](t_1)
\ldots
[Q(\lambda), a_l](t_l) \>Q'(\lambda) (t_{l+1})\>
[Q(\lambda), a_{l+1}](t_{l+2})\cr
&\hskip2in\ldots [Q(\lambda), a_{2k-1}](t_{2k}) \>e^{-H} \Bigl\}
d^{2k}t, \cr}
$$
for $a_0,\ldots, a_{2k-1} \in {\cal A}^1$. Observe that the heat
kernels in ${\rm Ch}(Q(\lambda))$ are independent of $\lambda$ and so
differentiating them is trivial. Differentiating the commutators
with $Q(\lambda)$ is done by means of \qprref . The arguments of
Proposition III.5 of [13] show that, algebraically,
$$
{d\over d\lambda} {\rm Ch}(Q(\lambda)) = (b+B)G^\lambda,
\num
$$
where $b$ and $B$ are the coboundary operators of entire cyclic cohomology.
We will be finished if we prove that each $G^\lambda_{2k-1}$ is well-defined
and obeys the growth condition of entire cyclic cohomology.

The key estimate on $Q(\lambda)$ is
$$
\bigl\Vert Q(\lambda) e^{-sH} \bigr\Vert \le Cs^{-1/2},  \num
$$
which follows from the spectral theorem. We proceed as in the derivation of
\jlo, by applying H\"older's inequality to the trace in \gdef.
We obtain the estimate
$$
\eqalign{&\Bigl|\Tr(A_0e^{-s_0 H}A_1e^{-s_1H}\ldots
Q'(\lambda) e^{-s_lH}
\ldots A_{2k-1} e^{-s_nH})\Bigr| \cr
&\qquad\le  \bigl\Vert Q'(\lambda) e^{-s_lH/2} \bigr\Vert
\>\Tr(e^{- H})^{1-s_l} \Tr(e^{- H/2})^{s_l}
\prod_{j=0}^{2k-1} \Vert A_j\Vert\cr
&\qquad\le  C {s_l}^{-1/2} \prod_{j=0}^{2k-1}
\Vert A_j\Vert\; .\cr}  \num
$$
Using this estimate, the integral over $\sigma^1_{2k}$ in \gdef\ is
well-defined and gives a factor of ${1\over (2k-1)!}$.  The sum in \gdef\
involves $2k$ such terms, so the resulting bound is
$$
\bigl| G^\lambda_{2k-1}(a_0,\ldots,a_{2k-1}) \bigr|
\le C\; {2k\over (2k-1)!} \; \Vert a_0 \Vert \prod_{l=1}^{2k-1}
\bigl\Vert [Q(\lambda), a_l] \bigr\Vert, \num
$$
where $C$ is independent of $k$.  This shows
that the growth condition on $G^\lambda$ is satisfied and so
$G^\lambda$ is an entire cochain. $\quad\square$

\subsec
In fact, formula \cocycle\ defines a one parameter family of cocycles
indexed by $\beta$ (we will refer to $\beta$ as the temperature because
of the obvious analogy with quantum statistical mechanics). It is shown
in [14] and [16] that the entire cyclic cohomology class of
${\rm Ch}^\beta(Q)$ is independent of $\beta$. It is thus natural
to study the limit $\beta\rightarrow\infty$. We have
the following theorem ([11], Section 2.2).

\thm\cmthm{The zero temperature limits,
$$
{\rm Ch}^\infty_{2k}(Q)(a_0,a_1,\dots ,a_{2k}):=
\lim_{\beta\rightarrow\infty}{\rm Ch}^\beta_{2k}(Q)(a_0,a_1,\dots ,
a_{2k}),\num
$$
exist and define continuous $(2k+1)$--linear functionals over
the algebra ${\cal A}^Q$. Moreover,
$$
{\rm Ch}^\infty_{2k}(Q)(a_0,a_1,\dots ,a_{2k})=
{{(-1)^k}\over{k!}}\Str \Bigl\{P_0a_0P_0\Omega(a_1,a_2)P_0\dots
P_0\Omega(a_{2k-1},a_{2k})\Bigr\},\num
$$
where $P_0$ is the orthogonal projection onto ${\rm Ker}(H)$, and
where $\Omega(a,b)=ab-aP_0b$.}

\medskip\noindent
In the examples studied in this paper, ${\rm Ker}(Q)$ is a one dimensional
subspace of the even part ${\cal H}_0$ of $\cal H$. The above theorem then
yields the following corollary.

\cor\zerothm{Let ${\rm Ker}(Q)$ be a one dimensional subspace of
${\cal H}_0$
and let $a_0,a_1,\dots ,a_{2k}\in{\cal A}^Q$. Then
$$
{\rm Ch}^{\infty}_{2k}(Q)(a_0,a_1,\dots ,a_{2k})=
{(-1)^k\over k!}\langle a_0\rangle_0\prod_{m=0}^k \Bigl\{
\langle a_{2m-1}a_{2m}\rangle_0-\langle a_{2m-1}\rangle_0
\langle a_{2m}\rangle_0\Bigr\} ,\ref{\zerolim}
$$
where $\langle a\rangle_0:=(\phi_0, a\phi_0)$, and where
$\phi_0$ is a normalized vector spanning ${\rm Ker}(Q)$.}

\medskip\noindent
{\it Remark}. The factors $\langle ab\rangle_0-\langle a\rangle_0
\langle b\rangle_0$ appearing in \zerolim\ are the
truncated vacuum expectation values of $a$ and $b$.

\medskip\noindent
{\it Proof.} As a consequence of \cmthm ,
$$
\eqalign{
{\rm Ch}^{\infty}_{2k}(Q)&(a_0,a_1,\dots ,a_{2k})=\cr
&{{(-1)^k}\over{k!}}(\phi_0,a_0P_0\Omega(a_1,a_2)P_0\dots
P_0\Omega(a_{2k-1},a_{2k})\phi_0)=\cr
&{{(-1)^k}\over{k!}}\langle a_0\rangle_0\prod_{m=0}^k \Bigl\{
\langle a_{2m-1}a_{2m}\rangle_0-\langle a_{2m-1}\rangle_0
\langle a_{2m}\rangle_0\Bigr\} ,\cr}
$$
as claimed. $\square$

Note, however, that the limit ${\rm Ch}^\infty(Q):=
\{{\rm Ch}^\infty_{2k}(Q)\}_{k=0}^\infty$ of ${\rm Ch}^\beta (Q)$
does not define an entire cyclic cocycle. The power series
$\sum_{k\geq 0}k!\; ||{\rm Ch}_{2k}^\infty(Q)||\; z^{2k}$ has a finite,
rather than infinite, convergence radius, and so Connes' growth
condition is violated. As a consequence, the usual pairing [10], [14],
$<{\rm Ch}^\infty (Q), e>$ of ${\rm Ch}^\infty (Q)$ with a $K_0({\cal A})$
class $e$ is meaningless. It is, however, easy to see that if a hermitian
projection $e\in{\rm Mat}({\cal A})$ is such that the operator
$I - 2P_0eP_0$ is invertible, then the series defining $<{\rm Ch}^\infty (Q),
e>$ converges and, in fact
\medskip
$$
<{\rm Ch}^\infty (Q), e> = {1\over 2}\Str_{{\rm Ker}(Q)}
\big\{I-{{I - 2P_0eP_0}\over{[(I - 2P_0eP_0)^2]^{1/2}}}\big\}\; . \num
$$
\medskip\noindent
Note that $<{\rm Ch}^\infty (Q), e>$ is an integer such that
$|<{\rm Ch}^\infty (Q), e>|\leq \dim {\rm Ker}(Q)$. We are not aware
of a topological significance of this integer.

\section\suisec{Fredholm Modules Over the Quantum Type I
Cartan Domains}

\newsubsec
The Cartan domains of type I form an infinite sequence $D_{m,n},
\; m,n\geq 1$, of non-compact hermitian symmetric spaces.
$D_{m,n}$ is an open subset of $\bC^{mn}$ defined as follows:
$$
D_{m,n}:=\{z\in {\rm Mat}_{m,n}(\bC):I_m-zz\star>0\}\; .\num
$$
The quantum deformation of $D_{m,n}$ is the Toeplitz algebra
$\trd$, defined as follows [7]. For $r>m+n-1$,
we consider the following measure on $D_{m,n}$:
$$
d\mu_r(z)=\Lambda_r\det (I_m-zz\star)^{r-(m+n)}d^{2mn}z\;
.\num
$$
The normalization factor $\Lambda_r$ is chosen to normalize the
total integral to one:
$$
\Lambda_r=\pi^{-mn}\prod_{k=1}^n{{\Gamma (r-n+k)}
\over{\Gamma (r-m-n+k)}}\; .\num
$$
We let $\hru$ denote the Hilbert space
of holomorphic functions on $D_{m,n}$ which are square integrable
with respect to $d\mu_r$. The Bergman kernel of $D_{m,n}$
associated with this measure is given
by $K^r(z,w)=\det (I_m-zw\star)^{-r}$. The algebra
$\trd$ is the $\bC\star$-algebra generated by the Toeplitz operators
on $\hru$ whose symbols are smooth functions on
$D_{m,n}$ which extend to the closure $\overline{D}_{m,n}$.
Its generators $\sigma_{ij}:=T_r(z_{ij})$ and ${\bar\sigma}_{ij}
=T_r({\bar z}_{ij})$ obey the relations
$$
\eqalign{
&[{\bar\sigma}_{ij},\sigma_{kl}]=\mu (I-\sigma\sigma\star)_{ki}
(I-\sigma\star\sigma)_{jl}\; ,\cr
&[\sigma_{ij},\sigma_{kl}]=0\; ,}\num
$$
where $\mu=1/(r-m)$.

\subsec
Our construction of Fredholm modules over $\trd$
will be based on a quantization of the type I Cartan superdomain
$\su\equiv D^I_{m,n|n}$ [6]. The starting point of this
construction is the
Hilbert space $\Hru$ of superholomorphic functions on $\su$ which
are square integrable with respect to the measure
$$
d\mu_r(Z)=
{1\over \pi^{mn}} \det(I_m-ZZ\star)^{r-m} d^{2mn}z\>
d^{2mn}\theta. \num
$$
Here, $\theta$ denotes the matrix of fermionic generators
and $Z=(z,\theta)$ is a collective matrix notation for the generators
of $C^\infty(\su)$.  The corresponding
Bergman kernel is given by
$$
\rker ZW = \det (I_m - ZW\star)^{-r}. \num
$$
Let $\tru$ denote the $\bC\star$-algebra generated by the super
Toeplitz operators on $\Hru$ with smooth symbols extending to the
boundary. Its generators are $\Sigma_{ij} = T_r(Z_{ij})$ and
${\bar\Sigma}_{ij}=T_r({\bar Z}_{ij})$. Note that $\bar\Sigma_{ij}$ is
the adjoint of $\Sigma_{ij}$.  We will adopt a matrix notation:
$\Sigma\star_{\phantom{\ast}ij} = \bar\Sigma_{ji}$. Often we
will write $\Sigma = (\sigma,\chi)$ to indicate the submatrices of
even and odd operators.

The theorem below applies to all Type I Cartan superdomains $\cald$,
not just the supersymmetric case that we have denoted by $\su$.
\thm\rltnthm{Using the above notation, the generators of ${\cal T}_r(\cald)$,
where $\cald = D^I_{m,n|q}$ is an arbitrary type I Cartan superdomain,
satisfy the following  relations:
$$
\eqalign{&[\Sigma_{ij}, \Sigma_{kl}] = 0, \cr
&[\bar\Sigma_{ij},\Sigma_{kl}] =  \mu(I_m-
\Sigma\Sigma\star)_{ki} (\inq - \Sigma\star\Sigma)_{jl},  \cr}
\num
$$
where $[\cdot,\cdot]$ is the graded commutator, and
$\mu = 1/(r - m)$.   In other words,
$$
\eqalign{&[\bar\sigma_{ij},\sigma_{kl}] = \mu(I_m-
\sigma\sigma\star-
\chi\chi\star)_{ki} (I_n - \sigma\star\sigma)_{jl},  \cr
&[\bar\sigma_{ij},\chi_{kl}] =  - \mu(I_m-\sigma\sigma\star -
\chi\chi\star)_{ki} (\sigma\star\chi)_{jl},  \cr
&[\bar\chi_{ij},\chi_{kl}] = \mu(I_m-\sigma\sigma\star -
\chi\chi\star)_{ki} (I_n - \chi\star\chi)_{jl}.  \cr}\num
$$}

\medskip\noindent
Before proving \rltnthm, we first prove two lemmas.  Let
$\Delta_{ij}$ be the unbounded operator on $\hil_r(\cald)$ given by
$$
\Delta_{ij} \phi(Z) = {\del\over \del Z_{ij}} \phi(Z) .  \num
$$

\lemma\issdlemma{For $\phi\in\hil_r(\cald)$ in the domain of
$\Delta_{ij}$ for all $1\le i\le m$, we have
$$
\sum_{j=1}^m \Delta_{jk}(I_m - \Sigma\Sigma\star)_{ji} \phi =
\mu\inv \bar\Sigma_{ik}\phi.  \ref{\issd}
$$
The domain of this operator thus extends to all of $\hil_r(\cald)$.}

\medskip\noindent{\it Proof.} By definition we have
$$
\sum_{j=1}^m \Delta_{jk}(I_m - \Sigma\Sigma\star)_{ji}\phi(Z) =
 \sum_{j=1}^m \int_\su {\del\over \del Z_{jk}}
\Bigl[\rker ZY (I_m - ZY\star)_{ji}\Bigr] \phi(Y)  \dmu Y .  \num
$$
To evaluate the derivative, we need the fact that
$$
{\del\over \del Z_{jk}} \rker ZY =   r \rker ZY
\bigl[Y\star(I_m - ZY\star)\inv \bigr]_{kj}.  \num
$$
We thus obtain
$$
\eqalign{&\sum_{j=1}^m {\del\over \del Z_{jk}} \Bigl[\rker ZY (I_m -
ZY\star)_{ji}\Bigr] \cr
&\qquad = \rker ZY \sum_{j=1}^m \Bigl[ r \bigl[Y\star(I_m -
ZY\star)\inv \bigr]_{kj} (I_m - ZY\star)_{ji} -  \bar Y_{ik} \Bigr]  \cr
&\qquad =  (r - m)  \bar Y_{ik} \rker ZY.  \cr}\num
$$
The lemma follows.  $\square$

\lemma\disslemma{For $\phi\in\hil_r(\cald)$ in the domain of $\Delta_{ij}$
for all $1\le j\le n+q$, we have
$$
\sum_{k=1}^{n+q}   (\inq - \Sigma\star\Sigma)_{jk} \Delta_{lk}\phi =
\mu\inv \bar\Sigma_{lj}\phi.  \ref{\diss}
$$
The domain of this operator thus extends to all of $\hil_r(\cald)$.}

\medskip\noindent{\it Proof.}
We start with
$$
\sum_{k=1}^{n+q}   (\inq - \Sigma\star\Sigma)_{jk} \Delta_{lk} \phi(Z) =
\sum_{k=1}^{n+q} \int_\su \rker ZW  (\inq - W\star W)_{jk} {\del\over \del
W_{lk}}\phi(W)  \dmu W. \num
$$
Integrating by parts gives
$$
\eqalignno{&\sum_{k=1}^{n+q} (\inq - \Sigma\star\Sigma)_{jk} \Delta_{lk}
\phi(Z) &\eqalignref{\iparts} \cr
&= - \sum_{k=1}^{n+q}  (-1)^{\epsilon_k(\epsilon_j +1)}\int_\su \rker ZW
{\del\over \del W_{lk}} \Bigl[(\inq - W\star W)_{jk} \det(\inq -
W\star W)^{r-m} \Bigr] \phi(W)  dW,\cr}
$$
where $\epsilon_j := p(Z_{ij})$.
The derivative is easily computed,
$$
\eqalign{&{\del\over\del W_{lk}} (\inq - W\star W)_{jk} \det(\inq -
W\star W)^{r-m} \cr
&\qquad=  - \det(\inq - W\star W)^{r-m}
\biggl[(-1)^{\epsilon_k\epsilon_j}\bar W_{lj} \cr
&\qquad\qquad+
(-1)^{\epsilon_k(\epsilon_j-1)} (r-m) (\inq - W\star W)_{jk}
\bigl[(\inq - W\star W)\inv W\star\bigr]_{kl}  \biggr] .\cr} \num
$$
Summing over $k$ we obtain
$$
\eqalign{&- \sum_{k=1}^{n+q}  (-1)^{\epsilon_k(\epsilon_j +1)}
{\del\over\del W_{lk}}
(\inq - W\star W)_{jk} \det(\inq - W\star W)^{r-m} \cr
&\qquad\qquad=  (r-m) \bar W_{lj} \det(\inq - W\star W)^{r-m}.\cr}\num
$$
In view of \iparts, this completes the proof. $\quad\square$

\bigskip\noindent{\it Proof of \rltnthm.} We start with the fact that
$$
[\Delta_{ab},\Sigma_{kl}] = \delta_{ak} \delta_{bl}, \num
$$
restricted to the domain of $\delta_{ij}$.
We apply operators to both sides of this equation and contract
indices:
$$
\sum_{a,b} (\inq - \Sigma\star\Sigma)_{jb}
[\Delta_{ab},\Sigma_{kl}]
(I_m - \Sigma\Sigma\star)_{ai} = (\inq - \Sigma\star\Sigma)_{lj}
(I_m - \Sigma\Sigma\star)_{ki}. \num
$$
Using \issdlemma\ and \disslemma, we reduce the right-hand side
to
$$
\mu\inv \sum_{a} \bar\Sigma_{aj} \Sigma_{kl} (I_m -
\Sigma\Sigma\star)_{ai} - \mu\inv \sum_{b} (-1)^{\epsilon_b
\epsilon_l} (\inq - \Sigma\star\Sigma)_{jb} \Sigma_{kl}
\bar\Sigma_{ib}   \mu\inv [\bar\Sigma_{ij}, \Sigma_{kl}].\num
$$
This proves the theorem on a restricted domain.  It is easy to see that
this domain is dense, and since both sides of the relation are bounded
operators,  there is no problem in removing the restriction.
$\quad\square$

\subsec
For $f\in C^{\infty}(D_{m,n})$ bounded, the Toeplitz operator
$T_r(f)$
defines a unique super Toeplitz operator which we will denote also
by $T_r(f)$. This defines an action of $T_r(f)$ on the $\bZ_2$-graded
Hilbert space $\Hru$. A continuity argument shows that this action
extends to an action of the $\bC\star$-algebra
$\trd$ on $\Hru$, and so we have a $\ast$-morphism $\rho :\trd
\longrightarrow {\cal L}(\Hru)$.

Let ${\cal P}\subset\hru$ denote the dense subspace spanned
by all polynomials. We now take ${\rm Dom}(\Delta_{ij})=
{\cal P}$. This operator is broken up into its even and odd
components as $\Delta = (\del,\tau)$, where, if $Z  = (z,\theta)$,
$$
\del_{ij}\phi(Z) = {\del\over \del z_{ij}}\phi(Z), \qquad
\tau_{ij}\phi(Z) = {\del\over \del \theta_{ij}} \phi(Z) . \num
$$
Let $\bar\Delta_{ij}, \bar\del_{ij},$ and $\bar\tau_{ij}$
denote the hermitian adjoints of $\Delta_{ij}, \del_{ij},$
and $\tau_{ij}$, respectively.   Now we define the operator
$$
d := \sum_{ij} \chi_{ij} \del_{ij}, \num
$$
and let $d\star := \sum_{ij} \bar\chi_{ij} \bar\del_{ij}$ denote its adjoint.

The two operators
$$
\eqalign{
&Q_1:=d + d\star,\cr
&Q_2 :=i(d - d\star),\cr}\ref{\supercharges}
$$
are defined on ${\cal P}$ and symmetric. Let $N_0$ and $N_1$
denote the operators on ${\cal P}$,
$$
\eqalign{&N_0 := \sum_{i,j} \sigma_{ij} \del_{ij}, \cr
&N_1 := \sum_{i,j} \chi_{ij} \tau_{ij}.  \cr} \num
$$
Note that on monomials these operators have the form
$$
\eqalign{
&N_0(z^\mu\theta^\alpha):=|\mu|z^\mu\theta^\alpha\; ,\cr
&N_1(z^\mu\theta^\alpha):=|\alpha| z^\mu\theta^\alpha\;
,\cr}\num
$$
where $\mu\in\bZ_+^{\phantom{+} mn}$ and $\alpha\in\{0,1\}^{
mn}$ are multi-indices and $|\mu| = \mu_1+\ldots+\mu_{mn}$.
Let $H$ be the total number operator,
$$
H:=N_0+N_1.\num
$$
Then $N_0$ is symmetric and $N_1$ is bounded and self-adjoint.
\prop\qsqprop{As operators on ${\cal P}$,
$$
Q_1^2=Q_2^2=H .\num
$$}
\medskip\noindent{\it Proof.}
Using \issdlemma\ we have
$$
\eqalign{d\star &=
\mu \sum_{i,j,k} \tau_{kj} (I_m - \Sigma\Sigma\star)_{ki}
\bar\del_{ij} \cr
&= \sum_{j,k} \sigma_{jk} \tau_{jk}  .\cr} \ref{\barchidel}
$$
Thus,
$$
Q_1 = \sum_{j,k} \chi_{jk} \del_{jk}
+ \sigma_{jk} \tau_{jk}, \num
$$
and
$$
\eqalign{Q_1^2 &=  \sum_{i,j,k,l}[\chi_{ij}\del_{ij}, \sigma_{kl}
\tau_{kl}]
\cr
&= \sum_{i,j,k,l} \chi_{ij} [\del_{ij}, \sigma_{kl}]
\tau_{kl}  +  \sigma_{kl} [\chi_{ij}, \tau_{kl}] \del_{ij}\cr
&= N_1 + N_0. \cr}  \num
$$
The proof for $Q_2$ is essentially identical.  $\quad\square$

\prop\saprop{
\item{(i)} The operators $Q_j$, $H$, and $N_0$ are essentially
self-adjoint on ${\cal P}$.
\item{(ii)} For any $\epsilon > 0$, $(H+I)^{-1/2}\in
I_{2mn+\epsilon}(\Hru)$.}
\medskip\noindent{\it Proof.}
(i) Let $\phi\in\calp$ be a polynomial of degree $m$.
Then $||H^k\phi|| \le Cm^k$, with $C$ independent of $k$.  As a
consequence,
each $\phi\in\calp$ is an analytic vector for $H$, and thus $H$ is
essentially
self-adjoint on $\calp$ by Nelson's theorem [18]. Since $N_0 = H-N_1$, with
$N_1$ bounded, the same is true for $N_0$.  Finally, $Q_j$ is essentially
self-adjoint on $\calp$ as $\norm{Q_j^k\phi} \le Cm^{k/2}$ for all
$\phi\in\calp$.

(ii) The spectrum of $N_0$ consists of the eigenvalues $\lambda_p
=p,\; p=0, 1, 2,\dots$ each of which has multiplicity not exceeding
$2^{mn}\times$ the number of monomials in $z_{ij}$ of degree $p$
$=O(p^{mn-1})$. Since $N_1$ is bounded, the claim follows.
$\quad\square$

\bigskip
The following proposition states that the operators $Q_1$
and $Q_2$ generate an $N=2$ supersymmetry algebra.

\prop\susyprop{As operators on ${\cal P}$, we have the
following relations:
$$
\eqalign{
&[Q_1,\sigma_{ij}]=\chi_{ij} ,\quad
[Q_1,\bar\sigma_{ij}]=-\bar\chi_{ij},\quad
[Q_1,\chi_{ij}]=-\sigma_{ij},\quad
[Q_1,\bar\chi_{ij}]=\bar\sigma_{ij},\cr
&[Q_2,\sigma_{ij}]=i\chi_{ij},\quad
[Q_2,\bar\sigma_{ij}]=i\bar\chi_{ij},\quad
[Q_2,\chi_{ij}]=i\sigma_{ij},\quad
[Q_2,\bar\chi_{ij}]=i\bar\sigma_{ij}.\cr}\ref{\susyone}
$$
Furthermore, the operators $Q_j$ satisfy the relations:
$$
[Q_j,Q_k]=2\delta_{jk}H\; .\ref{\susy}
$$}
\medskip\noindent{\it Proof.}
We have trivially the following relations:
$$
\eqalign{
&[d,\sigma_{ij}]=\chi_{ij},\cr
&[d,\chi_{ij}]=0,\cr
&[d,\bar\chi_{ij}]=\bar\sigma_{ij},\cr}\ref{\chisa}
$$
In addition, we see using the adjoint of \barchidel\ that
$$
[d, \bar\sigma_{ij}] = 0,  \ref{\chisb}
$$
and
$$
[d, \bar\chi_{ij}] = \bar\sigma_{ij}.  \ref{\chisc}
$$
The relations \susyone\ follow from \chisa, \chisb, and \chisc.

Among the relations
\susy\ , only $[Q_1,Q_2]=0$ needs to be established. This, however,
is an immediate consequence of \supercharges. $\quad\square$

\bigskip
As a consequence of the above considerations and of \homthm,
we obtain the following theorem.
\thm\modthm{
\item{(i)} The two triples $(\Hru ,\;\rho ,\; Q_j)$
define $2mn$--summable Fredholm modules over $\trd$.
\item{(ii)} The corresponding Chern characters define the same
cohomology  class in the entire cyclic cohomology of $\trd^1$.}

\section\suiisec{The Chern Characters}

\newsubsec
Let $\apoly$ be the subalgebra of $\trd$ consisting of polynomials in
the Toeplitz operators. Clearly, an element of $a\in\apoly$ has
a representation by an integral kernel:
$$
a\phi(Z) = \int_\su a(Z,W) \phi(W), \num
$$
where $a(Z,W)$ is an even function (with respect to the grading) which is
holomorphic in $Z$ and  depends smoothly on $W$.
In this section we derive explicit expressions for
${\rm Ch}^\beta_{2k}(Q_j)(a_0,\dots ,a_{2k})$, for $a_j \in
\apoly$, and give representations for these functionals in terms of
multiple Berezin integrals over $\su$. We also consider the special
case when the $a_j$ are Toeplitz operators.

For the remainder of this subsection, $f$ denotes a smooth function
on $D_{m,n}$ whose first derivatives are bounded.
\prop\commprop{For $f$ as above,
$$
\eqalign{
&\sum_{i=1}^m [\chi_{ij} \del_{ik},T_r(f)] =  \sum_{i=1}^m T_r\bigl(
{\del f\over \del z_{ik}}\bigr) \chi_{ij} \cr
&\sum_{i=1}^m [\bar\chi_{ij} \bar\del_{ik},T_r(f)] =
- \sum_{i=1}^m \bar\chi_{ij}
T_r\bigl({\del f\over\del\bar z_{ik}}\bigr) .\cr}  \ref{\commone}
$$}
\medskip\noindent{\it Proof.}
It is sufficient to prove the first of these identities
as the second one follows by taking the hermitian conjugate.
Consider the commutator:
$$
\eqalign{\sum_{i=1}^m [\chi_{ij} \del_{ik},T_r(f)] &=
\sum_{i=1}^m\int_\su \theta_{ij} {\del\over\del z_{ik}} \rker ZW f(w) \phi(W)
\dmu W\cr
&\quad+\sum_{i=1}^m \int_\su \rker ZW {\del f\over \del w_{ik}}(w)  \eta_{ij}
\phi(W) \dmu W \cr
&\quad + \sum_{i=1}^m \int_\su \rker ZW f(w) \eta_{ij} \phi(W) {\del\over
\del w_{ik}} \dmu W.  \cr} \ref{\cdtcomm}
$$
We use
$$
{\del\over \del w_{ik}} \log \det(I_m - WW\star)^{-m}
= m \sum_{l=1}^m (I_m - WW\star)\inv_{\phantom{-1}li} \bar w_{lk},
\ref{\weusea}
$$
to rewrite the third term of \cdtcomm\ as
$$
m \sum_{i,l} \int_\su \rker ZW f(w) \eta_{ij} \bar w_{lk}
(I_m - WW\star)\inv_{\phantom{-1}li} \phi(W) \dmu W.
\ref{\thrdtrm}
$$
We make the $\eta_{ij}$ into a derivative using
$$
{\del\over \del \bar\eta_{lj}} \log \det(I_m - WW\star)^{-m}
= -m \sum_{i=1}^m (I_m - WW\star)\inv_{\phantom{-1}li} \eta_{ij}.
\ref{\weuseb}
$$
The third term of \cdtcomm\ thus becomes
$$
-  (-1)^{p(\phi)} \sum_{l=1}^m \int_\su \rker ZW f(w)  \bar w_{lk}
\phi(W) {\del\over\del \bar\eta_{lj}} \dmu W,  \num
$$
where $p(\phi)$ is the parity of $\phi$, which appears because
the $\eta$ was moved past the $\phi$.
If we integrate by parts, this parity is cancelled and the third term
of \cdtcomm\ becomes
$$
\sum_{l=1}^m \int_\su {\del\over\del \bar\eta_{lj}} \rker ZW f(w)
\bar w_{lk} \phi(W)  \dmu W, \num
$$
where the derivative strikes only the kernel.
Using computations essentially identical to \weusea\ and \weuseb,
we find that
$$
\sum_{l=1}^m \bar w_{lk} {\del\over\del \eta_{lj}} \rker ZW
= - \sum_{i=1}^m \theta_{ij} {\del\over\del w_{ik}} \rker ZW.  \num
$$
Thus, the third term in \cdtcomm\ cancels the first term, and the
proposition follows.  $\quad\square$

\cor\qcommcor{With the above definitions,
$$
\eqalign{
[Q_1,T_r(f)]=  \sum_{i,j} T_r\bigl({\del f\over \del
z_{ij}}\bigr)\chi_{ij}
- \bar\chi_{ij} T_r\bigr({\del f\over \del\bar z_{ij}}\bigr)
\; ,\cr
[Q_2,T_r(f)]= \sum_{i,j} iT_r\bigl({\del f\over \del
z_{ij}}\bigr)\chi_{ij}
+ i\bar\chi_{ij} T_r\bigr({\del f\over \del\bar z_{ij}}\bigr)
\; .\cr}  \ref{\commut}
$$}

\cor\apclosed{There is an inclusion $\apoly \subset \trd^1$.}

\subsec
For $t\in\sigma^\beta_n$ and $a_0,\dots ,a_n\in \apoly$
bounded, we now consider the expression
$$
\Str \bigl\{a_0 a_1(t_1)\dots a_n(t_n) \> e^{-\beta H}
\bigr\} \; .\ref{\form}
$$
As a consequence of \jlo, the supertrace \form\ is well
defined. Our goal in this subsection is to express
it as a multiple integral over $\su$. The integral representation
given below has the flavor of a Feynman-Kac representation in
Euclidean field theory.
\prop\strprop{Under the above assumptions,
$$
\eqalign{
\Str &\bigl\{a_0 a_1(t_1)\dots a_n(t_n)\>e^{-\beta H}\bigr\}\cr
&= \int_{\su^{n+1}} a_0(e^{-(\beta -t_n)}Z_n,Z_0)
a_1(e^{-t_1}Z_0,Z_1) a_2(e^{-(t_2-t_1)}Z_1,Z_2)  \ldots \cr
&\qquad\ldots a_n(e^{-(t_n-t_{n-1})}Z_{n-1},Z_n)
d\mu_r(Z_0)\dots d\mu_r(Z_n)\; . \cr}\ref{\formrep}
$$}
\medskip\noindent{\it Proof.}
Using a basis of homogeneous polynomials, we can write \form\ as
$$
\sum_{\alpha} (-1)^{p(\phi_\alpha)}
\bigl(\phi_\alpha,\; a_0 a_1(t_1)
\dots a_n(t_n)\phi_\alpha\bigr) e^{-\beta({\rm deg}\;
\phi_\alpha)}\; .\ref{\trace}
$$
Now, for $t>0$ and a holomorphic function $\phi\in{\rm Ran}(e^{-
tH})$,
we clearly have
$$
e^{tH}\phi(Z) = \phi(e^tZ),  \num
$$
by the definition of $H$.  This, in turn, implies that
$$
a(t)\phi(Z)=\int_{\su} a(e^{-t}Z,W) \phi(e^tW)
d\mu_r(W)\; .\ref{\toeplt}
$$
Using this fact we can rewrite \trace\  as
$$
\eqalign{
&\sum_{\alpha}(-1)^{p(\phi_\alpha)} \int_{\su^{n+2}}
\ol{\phi_{\alpha}
(e^{-\beta}W)} a_0(W,Z_0) a_1(e^{-t_1}Z_0, Z_1)
a_2(e^{-(t_2-t_1)}Z_1,Z_2)\ldots\cr
&\qquad\ldots a_n(e^{-(t_n-t_{n-1})}Z_{n-
1},Z_n)\phi_{\alpha}(e^{t_n}Z_n)
\>\dmu W d\mu_r(Z_0)\dots d\mu_r(Z_n)\; \cr
&= \sum_{\alpha}\int_{\su^{n+2}} a_0(W,Z_0) a_1(e^{-
t_1}Z_0,Z_1)
a_2(e^{-(t_2-t_1)}Z_1,Z_2)\dots a_l(e^{-(t_n-t_{n-1})}Z_{n-1}, Z_n)\cr
&\qquad\times\phi_{\alpha}(e^{t_n}Z_n)\ol {\phi_{\alpha}
(e^{-\beta}W)} \dmu W d\mu_r(Z_0)\dots d\mu_r(Z_n)\; .
\cr}\ref{\sumppb}
$$
Because
$$
\rker ZW = \sum_{\alpha} \phi_{\alpha}(Z)
\ol{\phi_{\alpha}(W)},  \num
$$
\sumppb\ reduces to
$$
\eqalign{&\int_{\su^{n+2}} a_0(W,Z_0) a_1(e^{-t_1}Z_0,Z_1)
a_2(e^{-(t_2-t_1)}Z_1,Z_2)\dots a_n(e^{-(t_n-t_{n-1})}Z_{n-1},
Z_n)\cr
&\qquad\times \rker {e^{-(\beta-t_n)}Z_n}W
\dmu W d\mu_r(Z_0)\dots d\mu_r(Z_n)\; . \cr}\ref{\sumppb}
$$
We perform the $W$ integration, yielding \formrep.
$\quad\square$

\bigskip
We now give the explicit formula for the Chern character.  Here as in
Section \fredsec, Ch$_{2k}^\beta(Q_j)$ denotes the $2k$-th component
of the Chern character associated with the Dirac operator $Q_j$.

\thm\chchthm{Let $f_0, \ldots, f_{2k} \in \smooth{D_{m,n}}$ have
bounded first derivatives.  On $\smooth\su$ define the differential
operator
$$
\theta\cdot{\del\over\del z} :=  \sum_{ij} \theta_{ij}
{\del\over \del z_{ij}} .\num
$$
Then, for $j = 1,2$ we have the integral representation:
$$
\eqalign{
&{\rm Ch}_{2k}^\beta(Q_j)\bigl(T_r(f_0),\ldots,T_r(f_{2k})\bigr)\cr
&\quad= (-\beta)^{-k} \int_{I^\beta_{2k+1}} \int_{\su^{2k+1}} f_0(z_0)
\prod_{m=1}^{2k}\{ \theta_{m}
\cdot {\del\over\del z_{m}} f_{m}(z_{m}) \> +(-1)^j
\bar\theta_{m} \cdot {\del\over \del\bar z_{m}}f_{m}(z_{m})\}\cr
&\qquad\times\prod_{l=0}^{2k}\rker {e^{-s_l}Z_l}{Z_{l+1}}\; \dmu{Z_l} \;
\delta\bigl(\beta - \sum_{l=0}^{2k} s_l \bigr)\; d^{2k+1}s,\cr}\num
$$
where $Z_{2k+1} := Z_0$.}

\medskip\noindent{\it Proof.}
Start with the definition \cocycle.  The variables $t_j$
are replaced by $s_j := t_{j+1} - t_j$.
We use then the integral representation following directly from \strprop.
We apply then \qcommcor\ to conclude the proof. $\quad\square$

\section\svsec{Fredholm Modules Over Quantized Vector Spaces}

\newsubsec
A complex vector space $V \cong \bC^n$ has a quantum
deformation given by
a Toeplitz $\bC^\ast$-algebra (see [8] and references therein).
The perturbed measure on $V$ is defined by
$\dmu z := {r^n\over \pi^n} \exp (-rz\cdot\bar z) d^{2n}z$ on $V$,
for $r > 0$, where $z\cdot\bar z = \sum_j z_j \bar z_j$. Let
$\hrv$ be the Hilbert space of holomorphic functions on $V$ which are
square integrable with respect to $d\mu_r$. The Bergman kernel for
$d\mu_r$ on $V$ is
$$
\rkerv zw = \exp(r z\cdot\bar w).  \num
$$
The Toeplitz algebra $\trv$ is the $\bC^\ast$-algebra generated by
the Toeplitz operators on $\hrv$ whose symbols are smooth bounded
functions on $V$.  Its ``generators''
$\sigma_i = T_r(z_i)$ and $\bar\sigma_i = T_r(\bar z_i)$, $1\le i
\le n$ obey the relations
$$
[\bar\sigma_j, \sigma_k] = {1\over r} \delta_{jk}.  \num
$$
Note that $\sigma$ and $\bar\sigma$ are not bounded,
so the algebra $\trv$ will be generated only
by certain bounded functions of these operators. This issue will
not be important here.

To construct Fredholm modules over $\trv$ we proceed as in Section
\suisec.The supervector space $\sv \cong \bC^{n|n}$ is the
supersymmetric version of $V$.  The quantum deformation of $\sv$
[5] is based on the Hilbert space $\hrsv$ of superholomorphic
functions on $\sv$ which are square integrable with respect to the
measure $\dmu Z := {1\over \pi} \exp(-rZ\cdot\bar Z) d^{2n}z \>
d^{2n}\theta$.  Here $\theta_i$, $1 \le i \le n$ denote the fermionic
generators, $Z = (z,\theta)$, and $Z\cdot\bar Z = \sum_j (z_j\bar
z_j + \theta_j\bar\theta_j)$.  The Bergman kernel for the measure
$d\mu_r$ on $\sv$ is
$$
\rker ZW = \exp(rZ\cdot\bar W).  \num
$$
We denote the algebra generated by super Toeplitz operators on
$\hrsv$ with smooth bounded
symbols on $\sv$ by $\trsv$. We define ``generators'' $\sigma_i := T_r(z_i)$,
$\bar\sigma_i := T_r(\bar z_i)$, $\chi_i := T_r(\theta_i)$ and
$\bar\chi_i := T_r(\bar\theta_i)$.  The operators $\sigma_i$ and
$\bar\sigma_i$ are not elements of $\trsv$, and they will be interpreted
as unbounded operators on $\hrsv$.  The generators satisfy the relations:
$$
\eqalign{
&[\bar\sigma_j, \sigma_k] = {1\over r} \delta_{jk}, \cr
&[\bar\chi_j, \sigma_k] = 0, \cr
&[\bar\chi_j, \chi_k] = {1\over r} \delta_{jk}, \cr
&[\sigma_j, \sigma_k] = [\chi_j, \chi_k]
= [\sigma_j, \chi_k] = 0, \cr}  \num
$$
\vfill\eject
\noindent
and their hermitian conjugates.

\subsec
We can decompose the Hilbert space $\hrsv$ into odd and even
subspaces, which are orthogonal.
As in Section \suisec, we can define a grading preserving
$\ast$-morphism $\rho :\trv \to {\cal L}(\hrsv)$.
We again let ${\cal P} \subset \hrsv$ denote
the dense subspace spanned by polynomials in $z$ and $\theta$.
For $1\le i \le n$ we define the operator
$$
\del_i \phi(Z) = {\del\over \del z_i} \phi(Z), \num
$$
with Dom$(\del_i) =  {\cal P}$, and let
$\bar\del_i$ denote its hermitian adjoint.  Let
$$
d := \sum_j \chi_j \del_j, \num
$$
with adjoint $d\star$.  Then the two operators
$$
\eqalign{&Q_1 := d+d\star,  \cr
&Q_2 := i(d - d\star),  \cr}  \num
$$
are defined on $\calp$ and symmetric.  Let $N_0$ and $N_1$ denote
the following operators on $\calp$:
$$
\eqalign{&N_0(z^\mu\theta^\alpha) := |\mu| \>z^\mu
\theta^\alpha, \cr
&N_1(z^\mu\theta^\alpha) := |\alpha| \>z^\mu \theta^\alpha, \cr} \num
$$
where $\mu$ and $\alpha$ are multi-indices.
$N_0$ is symmetric, $N_1$ is bounded and self-adjoint, and we let
$H := N_0 + N_1$.

\prop\qsqprop{As operators on $\calp$,
$$
Q_1^2 = Q_2^2 = H.  \num
$$
In other words, $H$ is the Laplace operator corresponding to both
$Q_1$ and $Q_2$.}

\medskip\noindent{\it Proof.}
We make use of the orthonormal basis for $\hrsv$ [5],
$$
\phi_{\mu,\alpha}(Z) := \biggl({r^{|\mu| + |\alpha|} \over \mu!}
\biggr)^{1/2} z^\mu \theta^\alpha,  \ref{\vbasis}
$$
where $\mu! = \mu_1! \ldots \mu_n!$ and $\theta^\alpha$ is
ordered  $\theta_1^{\alpha_1}\ldots \theta_n^{\alpha_n}$.  We easily derive
$$
\eqalign{
&\sigma_j\phi_{\mu,\alpha} = [(\mu_j+1)/ r]^{1/2} \phi_{\mu +
1_j, \alpha}, \cr
&\bar\sigma_j\phi_{\mu,\alpha} = (\mu_j/ r)^{1/2} \phi_{\mu -
1_j, \alpha}, \cr
&\chi_j\phi_{\mu,\alpha} = (-1)^{\{\sum_{k<j} \alpha_k\}}\>
(1 - \alpha_j) r^{-1/2} \phi_{\mu, \alpha + 1_j}, \cr
&\bar\chi_j\phi_{\mu,\alpha} = (-1)^{\{\sum_{k<j} \alpha_k\}}\>
\alpha_j r^{-1/2} \phi_{\mu, \alpha - 1_j}, \cr
&\del_j\phi_{\mu,\alpha} = (\mu_j r)^{1/2} \phi_{\mu - 1_j,
\alpha}, \cr
&\bar\del_j\phi_{\mu,\alpha} = [(\mu_j+1) r]^{1/2} \phi_{\mu +
1_j, \alpha}, \cr} \ref{\expop}
$$
where $1_j$ is the multi-index with 1 in the $j$-th place and zeroes
elsewhere.  We compute using \expop:
$$
\eqalign{Q_j^2 &= [d,d\star] \cr
&=   \sum_{k,l} \Bigl\{  \chi_l \bar\chi_k [\del_l, \bar\del_k]
+  [\chi_l, \bar\chi_k] \bar\del_k \del_l \Bigr\} \cr
&= \sum_k \Bigl\{r \chi_k \bar\chi_k + {1\over r} \bar\del_k
\del_k \Bigr\}\cr
&= N_1 + N_0, \qquad\square \cr}  \num
$$
for $j=1,2$.

\prop\qhnprop{
\item{(i)} The operators $Q_j$, $H$ and $N_0$ are essentially
self-adjoint on $\calp$.
\item{(ii)} For any $\epsilon > 0$, $(H+I)^{-1/2} \in
I_{2n+\epsilon}(\hrsv)$.}

\medskip\noindent{\it Proof.}
The proof (i) follows that of \saprop\ (i).  For (ii) we observe that the
norm
$$
\eqalign{\bigl\Vert (H+I)^{-1/2} \bigr\Vert_{2l+\epsilon}^2
&= \sum_{\mu,\alpha} \bigl(1 + |\mu| + |\alpha|\bigr)^{-l-
\epsilon/2} \cr
&\le n\sum_\mu (1 + |\mu|)^{-l-\epsilon/2}, \cr} \num
$$
is finite for all $\epsilon>0$ precisely when $l=n$.  $\quad\square$

\bigskip
The following proposition states that the operators $Q_j$ with
$j= 1,2$ generate an $N=2$ supersymmetry algebra.
\prop\susypropv{As operators on $\calp$, we have the following
relations
$$
\eqalign{&[Q_1, \sigma_j] = \chi_j, \quad [Q_1, \bar\sigma_j] =
-\bar\chi_j, \quad
[Q_1, \chi_j] = - \sigma_j, \quad [Q_1, \bar\chi_j] = \bar\sigma_j,
\cr
&[Q_2, \sigma_j] = i\chi_j, \quad [Q_2, \bar\sigma_j] =
i\bar\chi_j, \quad
[Q_2, \chi_j] = i\sigma_j, \quad [Q_2, \bar\chi_j] = i\bar\sigma_j.
\cr}\num
$$
Furthermore, the operators $Q_j$ satisfy
$$
[Q_j, Q_k] = 2\delta_{jk} H.  \num
$$}

\medskip\noindent{\it Proof.}
These are easily derived from \expop.  $\quad\square$

\bigskip\noindent
Following the arguments of \modthm\ we establish the following result.
\thm\modthmv{
\item{(i)} The two triples $(\hrsv ,\;\rho ,\; Q_j)$
define $2n$-summable Fredholm modules over $\trv$.
\item{(ii)} The corresponding Chern characters define the same
cohomology class in the entire cyclic cohomology of $\trv^1$.}

\subsec
We now proceed to find explicit expressions for the Chern
characters,
following the approach of Section~\suiisec.
\prop\commpropv{For $f \in C^\infty(V)$ bounded with bounded
first derivative, we have the relations
$$
\eqalign{&[\del_j, T_r(f)] = T_r\bigl({\del\over\del z_j} f), \cr
&[\chi_j, T_r(f)] = 0, \cr} \num
$$
and their complex conjugates.}

\medskip\noindent{\it Proof.}
The second property follows immediately from the factorization
$$
\dmu Z = \exp(-rz\cdot\bar z) \>\exp(-r\theta\cdot\bar\theta)\>
d^{2n}z \> d^{2n}\theta. \num
$$
To prove the first we evaluate
$$
\eqalign{\del_j T_r(f) \phi(Z) &= \int_\sv {\del\over\del z_j}
\rker ZW f(w) \phi(W) \dmu W \cr
&= r \int_\sv \bar w_j \rker ZW f(w) \phi(W) \dmu W. \cr}
\num
$$
We also have
$$
\eqalign{T_r(f) \del_j \phi(Z) &= \int_\sv
\rker ZW f(w) {\del\over \del w_j} \phi(W) \dmu W \cr
&= r \int_\sv \rker ZW f(w) \phi(W) \bar w_j \dmu W \cr
&\qquad - \int_\sv \rker ZW {\del\over\del w_j} f(w) \phi(W)
\dmu W, \cr}  \num
$$
using an integration by parts.  This completes the proof.
$\quad\square$

\cor\qcommcor{For $f \in C^\infty(V)$ bounded with bounded first
derivatives, we have the relations
$$
\eqalign{&[Q_1, T_r(f)] = \sum_j T_r\Bigl({\del f\over\del z_j}
\theta_j - {\del f \over \del z_j} \bar\theta_j\Bigr), \cr
&[Q_2, T_r(f)] = i \sum_j T_r\Bigl({\del f\over\del z_j}  \theta_j +
{\del f \over \del z_j} \bar\theta_j\Bigr). \cr}\num
$$}

\subsec
For $t\in \sigma_m^\beta$ and $a_0,\ldots,a_m \in \apoly$,
we form the supertrace
$$
\Str \bigl\{ a_0 a_1(t_1) \dots a_m(t_m)\> e^{-\beta H}
\bigr\},  \ref{\formv}
$$
which is well-defined because of \jlo.
\prop\strpropv{Under the above assumptions,
$$
\eqalign{
\Str &\bigl\{a_0 a_1(t_1)\dots a_n(t_n)\>e^{-\beta H}\bigr\}\cr
&= \int_{\sv^{n+1}} a_0(e^{-t_1}Z_0,Z_1) a(e^{-(t_2-t_1)}Z_1,Z_2)
\dots a(e^{-(t_n-t_{n-1})}Z_{n-1},Z_n)\cr
&\qquad\times \rker{e^{-(\beta -t_n)}Z_n}{Z_0}
d\mu_r(Z_0)\dots d\mu_r(Z_n)\; . \cr}\ref{\formrepv}
$$}
\medskip\noindent{\it Proof.}
Using the basis \vbasis, we can write \formv\ as
$$
\sum_{\mu,\alpha} (-1)^{|\alpha|} \Bigl( \phi_{\mu,\alpha},
\>a_0 a_1(t_1) \dots a_m(t_m) \phi_{\mu,\alpha}
\Bigr) e^{-\beta(|\mu| + |\alpha|)} .  \num
$$
As the basis again consists of monomials, the proof follows that of
\strprop. $\quad\square$

\bigskip
At this point it becomes straightforward to evaluate the Chern
character associated to $Q_j$ on general elements of $\apoly$.
However, as this essentially repeats the statement of \chchthm,  we
will not write the result.  We confine ourselves to writing the integral
representation for the functional evaluated on Toeplitz operators.

\thm\chthmv{Let $f_0,\ldots,f_{2k} \in \smooth V$ have bounded
first derivatives.  On $\smooth\su$ define the differential
operator
$$
\theta\cdot{\del\over\del z} :=  \sum_{i} \theta_{i}
{\del\over \del z_i} .\num
$$
Then, for $j=1,2$,
$$
\eqalignno{
&{\rm Ch}^\beta_{2k}(Q_j)(T_r(f_0),\dots ,T_r(f_{2k}))\cr
&\quad= (-1)^{(j-1)k}\beta^{-
k}\int_{I^\beta_{2k+1}}\int_{\sv^{2k+1}}
\delta\bigl(\sum_{n=0}^{2k}s_n-\beta\bigr)
\prod_{l=0}^{2k} \rker{e^{-s_l}Z_l}{Z_{l+1}}
&\eqalignref{\chernv}\cr
&\qquad\times f_0(z_0) \prod_{m=1}^{2k}
\biggl\{\theta_m\cdot {\del\over \del z_m}f_m(z_m)+(-1)^j
{\bar\theta}_m\cdot {\del\over\del\bar z_m}f_m(z_m)\biggr\}
\prod_{n=0}^{2k}d\mu_r(Z_n)
\;d^{2k+1}s\; ,\cr}
$$
where $Z_{2k+1} := Z_0$.}

\vfill\eject

\centerline{\bf References}
\baselineskip=12pt
\frenchspacing

\bigskip

\item{1.}
Berezin, F.A.: General concept of quantization, {\it Comm. Math.
Phys.}
{\bf40} (1975), 153--174.

\item{2.}
Berezin, F.A.: Introduction to Superanalysis, D. Reidel Publ. Co.,
Dordrecht (1987).

\item{3.}
Born, M., and Jordan, P.: Zur Quantenmechanik, {\it Z. Physik},
{\bf 34} (1925), 858--888.

\item{4.}
Born, M., Heisenberg, W., and Jordan, P.: Zur Quantenmechanik II,
{\it Z. Physik}, {\bf 35} (1926), 557--615.

\item{5.}
Borthwick, D., Klimek, S., Lesniewski, A., and Rinaldi, M.:
Super Toeplitz operators and non-perturbative deformation
quantization of supermanifolds, {\it Comm. Math. Phys.},
{\bf 153} (1993), 49--76.

\item{6.}
Borthwick, D., Klimek, S., Lesniewski, A., and Rinaldi, M.:
Cartan superdomains, Super Toeplitz operators and deformation
quantization, to appear.

\item{7.}
Borthwick, D., Lesniewski, A., and Upmeier, H.:
Non-perturbative deformation quantization of Cartan domains,
{\it J. Funct. Anal.}, {\bf 113} (1993), 153--176.

\item{8.}
Coburn, L.A.: Deformation estimates for the Berezin-Toeplitz
quantization,
{\it Comm. Math. Phys.}, {\bf 149} (1992), 415--424.

\item{9.}
Connes, A.: Non-commutative differential geometry, {\it Publ.
Math. IHES},
{\bf 62} (1986), 94--144.

\item{10.}
Connes, A.: Entire cyclic cohomology of Banach algebras and
characters of $\theta$-summable Fredholm modules, {\it K-Theory},
{\bf 1} (1988), 519--548.

\item{11.}
Connes, A. and Moscovici, H.: Transgression and the Chern
character
in non-commu-\break tative $K$-homology, {\it Comm. Math.
Phys.}, to appear.

\item{12.}
Dirac, P.A.M.: The fundamental equations of quantum mechanics,
{\it Proc. Roy. Soc.}, {\bf A109} (1926), 642--653.

\item{13.}
Ernst, K., Feng, P., Jaffe, A., and Lesniewski, A.: Quantum K-theory,
II. Homotopy invariance of the Chern character, {\it J. Funct. Anal.},
{\bf 90} (1990), 355--368.

\item{14.}
Getzler, E., and Szenes, A.: On the Chern character of a theta-summable
Fredholm module, {\it J. Funct. Anal.}, {\bf 84} (1989), 343--357.

\item{15.}
Heisenberg, W.: \"Uber quantentheoretische Umdeutung kinematischer
und mechanischer Beziehungen, {\it Z. Physik}, {\bf 33} (1925),
879--893.

\item{16.}
Jaffe, A., Lesniewski, A., and Osterwalder, K.: Quantum K-theory,
I. The Chern character, {\it Comm. Math. Phys.}, {\bf 118} (1988),
1--14.

\item{17.}
Klimek, S., and Lesniewski, A.: Quantum Riemann surfaces, I. The
unit disc, {\it Comm. Math. Phys.}, {\bf 146} (1992), 103--122.

\item{18.}
Reed, M., and Simon, B.: Methods of Modern Mathematical Physics, II:
Fourier Analysis, Self-Adjointness, Academic Press, New York, San
Francisco, London (1975).

\item{19.}
Rieffel, M. A.: Deformation quantization of Heisenberg manifolds,
{\it Comm. Math.\break Phys.}, {\bf 122} (1989), 531--562.

\item{20.}
Rieffel, M. A.: Non-commutative tori --- a case study of
non-commutative differentiable manifolds, {\it Cont. Math.},
{\bf 105} (1990), 191--211.

\item{21.}
Witten, E.: Constraints on supersymmetry breaking, {\it Nucl. Phys.},
{\bf B202} (1982), 253--316.

\vfill\eject\end